\documentclass[12pt]{article}
\usepackage{amssymb}
\usepackage{graphicx}
\usepackage[cp866]{inputenc}

\begin{document}
 \begin{center}

{\bf Enhancement of low-mass dileptons in ultraperipheral collisions}

\vspace{2mm}

I.M. Dremin$^{a,}\footnote{dremin@lpi.ru}$,
S.R. Gevorkyan$^{b,}\footnote{gevs@jinr.ru}$,
D.T. Madigozhin$^{b,}\footnote{madigo@mail.ru}$\\

$^a${\it Lebedev Physical Institute, Moscow, Russia}

$^b${\it Joint Institute of Nuclear Research, Dubna, Russia}
\end{center}

Keywords: spectra, ion, ultraperipheral, collider, Universe

\vspace{1mm}

\begin{abstract}
It is shown that production of low-mass $e^+e^-$-pairs in ultraperipheral
nuclear collisions is enhanced due to the Sommerfeld-Gamow-Sakharov (SGS)
factor. This effect is especially strong near the threshold of creation
of unbound $e^+e^-$-pairs with low masses in the two-photon fusion.
Coulomb attraction of the non-relativistic components of such pairs
may lead to the increased intensity of 511 keV photons. It can be
recorded at the NICA collider and has some astrophysical implications.
The analogous effect can be observed at LHC in dilepton production.
\end{abstract}

\vspace{1mm}

PACS: 25.75.-q, 34.50.-s, 12.20.-m, 95.30.Cq \\

\vspace{1mm}

Production of $e^+e^-$-pairs in electromagnetic fields of colliding heavy ions
was first considered by Landau and Lifshitz in 1934 \cite{lali}. It was
shown that the total cross section of this process rapidly increases with
increasing energy $E$ as $\ln ^3E$ in asymptotics. This is still the
strongest energy dependence in particle physics. Moreover, the numerical factor 
$Z^4\alpha ^4$ compensates in the total cross section the effect of the small 
electromagnetic coupling $\alpha $ for heavy ions with large charge $Ze$. 
Therefore, the ultraperipheral production of $e^+e^-$-pairs (as well as 
$\mu ^+\mu ^-$ etc.) in ion collisions can become
the dominant mechanism at very high energies. It is already widely studied
at colliders. The heuristic knowledge of these processes is helpful in
understanding some astrophysical phenomena as well.

Abundant creation of pairs with rather low masses is the typical feature of
ultraperipheral interactions \cite{20dr}. Dileptons are produced in grazing 
collisions of interacting ions where two photons from their electromagnetic 
clouds interact and a lepton pair is created. Two-photon fusion production of 
lepton pairs has been calculated with both the equivalent photon approximation 
proposed in \cite{wei, wil} and via full lowest-order QED calculations
\cite{rac, bgms, sergey1, sergey2} reviewed recently in \cite{drufn}.
According to the equivalent photon approximation, the spectra of dileptons
created in ultraperipheral collisions
can be obtained from the general expression for the total cross section
\begin{equation}
\sigma _{up}(X)=
\int dx_1dx_2\frac {dn}{dx_1}\frac {dn}{dx_2}\sigma _{\gamma \gamma }(X).
\label{e2}
\end{equation}
Feynman diagrams of
ultraperipheral processes contain the subgraphs of two-photon interactions
leading to production of some final states $X$ (e.g., $e^+e^-$ pairs).
These blobs can be represented by the cross sections of these processes.
Therefore, $\sigma _{\gamma \gamma }(X)$ in (\ref{e2}) denotes the total cross
section of production of the state $X$ by two photons from the electromagnetic
clouds surrounding colliding ions and $dn/dx_i$ describe the densities of
photons carrying the share $x_i$ of the ion energy.

The~distribution of equivalent photons with a fraction of the nucleon energy
$x$ generated by a moving nucleus with the charge $Ze$ can be denoted as
\begin{equation}
\frac {dn}{dx}=\frac {2Z^2\alpha }{\pi x}\ln \frac {u(Z)}{x}
\label{flux}
\end{equation}
if integrated over the transverse momentum up to some value (see, e.g.,
\cite{blp}). The physical meaning of the ultraperipherality
parameter $u(Z)$ is the ratio of the maximum adoptable transverse momentum
to the nucleon mass as the only massless parameter of the problem.
Its value is determined by the form factors of colliding ions
(see, e.g., \cite{vyzh}).
It is clearly seen from Eq. (\ref{flux}) that soft photons with small
fractions $x$ of the nucleon energy dominate in these fluxes.

The cross section $\sigma _{\gamma \gamma }(X)$ usually inserted
in (\ref{e2}) in case of creation of the unbound dielectrons $X=e^+e^-$
is calculated in the lowest order perturbative approach and
looks \cite{blp,brwh} as
\begin{equation}
\sigma _{\gamma \gamma }(X)=\frac {2\pi \alpha ^2}{M^2}
[(3-v^4)\ln \frac {1+v}{1-v}-2v(2-v^2)],
\label{mM}
\end{equation}
where $v=\sqrt {1-\frac {4m^2}{M^2}}$ is the velocity of the pair components 
in the pair rest system, $m$ and $M$ are the electron and dielectron masses,
correspondingly. The cross section tends to 0 at the threshold of pair
production $M=2m$ and decreases as $\frac {1}{M^2}\ln M$ at very large $M$.

The distribution of masses $M$ of dielectrons is obtained after inserting
Eqs (\ref{flux}), (\ref{mM}) into (\ref{e2}) and leaving free one integration
there. One gets \cite{20dr}
\begin{equation}
\frac {d\sigma }{dM}=\frac {128 (Z\alpha )^4}{3\pi M^3}
[(1+\frac {4m^2}{M^2}-\frac {8m^4}{M^4})
\ln \frac {1+\sqrt {1-\frac {4m^2}{M^2}}}{1-\sqrt {1-\frac {4m^2}{M^2}}}-
(1+\frac {4m^2}{M^2})\sqrt {1-\frac {4m^2}{M^2}}]
\ln ^3\frac {u\sqrt {s_{nn}}}{M},
\label{sM}
\end{equation}
where $\sqrt {s_{nn}}$ is the c.m.s. energy per a nucleon pair.

The perturbative expression for the cross section $\sigma _{\gamma \gamma }(X)$
of (\ref{mM})  can be generalized to include the non-perturbative effects
crucial near the pair production threshold $M=2m$. It happens to be possible for Coulomb
interaction governing the behavior of the components of a pair.
At the production point, the components of pairs with low masses close to 2$m$
move very slowly relative to one another. They are strongly influenced by
the attractive Coulomb forces. In the non-relativistic limit, these
states are transformed by mutual interactions of the components to effectively
form a composite state whose wave function is a solution of the relevant
Schroedinger equation. The normalization of Coulomb wave functions plays an
especially important role at low velocities. It differs from the normalization
of free motion wave functions used in the perturbative derivation of
Eq. (\ref{mM}). 

 The amplitude $R_C$ of the process $\gamma\gamma\to e^+e^-$ with account of 
the interaction between leptons is connected to the amplitude $R_0$ without the
final state interaction by the relation 
\begin{equation}
R_C=\int \Psi_f(r)R_0(r)d^3r
\label{rc}
\end{equation}
where $\Psi_f(r)$ is the wave function for bound (parapositronium) 
or unbound lepton pairs in the coordinate representation. 

For lepton pairs in $S$-state (the orbital momentum $l$=0) the
characteristic distances of the pair production are $~1/m$,  whereas the 
Coulomb interaction between leptons acts over the much larger distances 
($~1/{m\alpha}$ in the bound state production and $~1/k$ for the unbound 
states ($k$ is the relative momentum). Therefore, the wave function can be
considered as constant in (\ref{rc}) and one gets
\begin{equation}
 R_C= \Psi_{kS}(r=0)\int R_0(r)d^3r= \Psi_{kS}(r=0) R_0(p=0).
 \label{rc0}
\end{equation}
This relation is valid not only for bound states, but also 
for the creation of the unbound lepton pairs if $kr_s\ll 1$. 
Such factorization of matrix elements has been widely used in the dimesoatoms 
production \cite{sergey3}. It is useful  for any process where the  
characteristic distances of pair production and of final state interactions 
are substantially different.
The normalization of the unbound pair wave function reads \cite{llqm}
\begin{equation}
|\psi_{kS}(\vec r=0)|^2 =\frac{\pi\xi}{sh(\pi\xi)}e^{\pi\xi}
=\frac{2\pi\xi}{1-e^{-2\pi\xi}};~~~\xi=\frac{2\pi\alpha m}{k}.
\label{psi}
\end{equation}
This is the widely used Sommerfeld-Gamow-Sakharov (SGS) factor 
\cite{som, gam, somm, sakh} which unites the non-perturbative and perturbative 
matrix elements. It results in the so-called "$\frac {1}{v}$-law" of the 
enlarged outcome of the reactions with extremely low-mass pairs produced. 
This factor is described in the standard
textbooks on non-relativistic quantum mechanics (see, e.g., \cite{llqm})
and used in various publications (e.g., \cite{baier, ieng, cass, arko}). 
The Sakharov recipe of its account for production of $e^+e^-$-pairs desctibed 
in \cite{sakh} consists in direct multiplication of the differential 
distribution of Eq. (\ref{sM}) by the SGS-factor written as
\begin{equation}
T=\frac {2\pi \alpha}{v(1-\exp (-2\pi \alpha/v))}.
\label{sgs}
\end{equation}
It enhances the contribution of the low-mass (low-$v$) pairs.
Thus the proper distribution of dielectron masses in ultraperipheral processes
is
\begin{eqnarray}
\frac {d\sigma }{dM^2}&=&\frac {128 (Z\alpha )^4}{3M^4}
\frac {\alpha }{\sqrt {1-\frac {4m^2}{M^2}}
(1-\exp (-2\pi \alpha/\sqrt {1-\frac {4m^2}{M^2}}))}\times\nonumber \\
& &\left[(1+\frac {4m^2}{M^2}-\frac {8m^4}{M^4})
\ln \frac {1+\sqrt {1-\frac {4m^2}{M^2}}}{1-\sqrt {1-\frac {4m^2}{M^2}}}-
(1+\frac {4m^2}{M^2})\sqrt {1-\frac {4m^2}{M^2}}\right]
\ln ^3\frac {u\sqrt {s_{nn}}}{M}, \nonumber \\
\label{sM2}
\end{eqnarray}
The distribution of the relative (in $e^+e^-$ rest system) velocity $v$ is like
\begin{equation}
\frac {d\sigma }{dv^2}=\frac {16(Z\alpha )^4}{3m^2}
[(3-v^4)\ln \frac {1+v}{1-v}-2v(2-v^2)]
\frac {\alpha}{v(1-\exp (\frac {-2\pi \alpha}{v}))}
\ln^3 \frac {u\sqrt {s_{nn}(1-v^2)}}{2m}.
\label{sv2}
\end{equation}
Let us remind that the velocity $v$ is related to the velocity of the
positron $v_+$ in the electron rest system as
\begin{equation}
v^2=\frac {1-\sqrt {1-v_+^2}}{1+\sqrt {1-v_+^2}}
\label{vv}
\end{equation}
so that $v_+=2v$ at $v\rightarrow 0$ and both $v$ and $v_+$ tend to 1
in the ultrarelativistic limit. The relative velocities $v$ and $v_+$ are
the relativistic invariants represented by the Lorentz-invariant
masses $m$ and $M$.

It has been shown in Ref. \cite{20dr} (see Fig. 1 there) that the cross
section of creation of unbound $e^+e^-$-pairs tends to zero at the threshold
$M=2m$ if the perturbative expression Eq. (\ref{mM}) is used. Account of the
non-perturbative SGS-factor (\ref{sgs}) in Eqs (\ref{sM2}) and (\ref{sv2})
drastically changes the situation, especially at low masses $M$ and
velocities $v$.

In Figs 1 and 2, we compare the yields of pairs with (curves a) and
without (curves b) account of the SGS-factor at NICA energy 11 GeV as
functions of masses $M$ and velocities $v$. It is clearly seen that the
overall contribution due to the correction is not high. It amounts to
about 4.6 percents at the peak of the $M^2$ distribution and 2.5 percents
at the peak of the $v^2$ distribution. The integral contributions differ 
by 3.4 percents only.

\begin{figure}
\includegraphics[width=\textwidth, height=8cm]{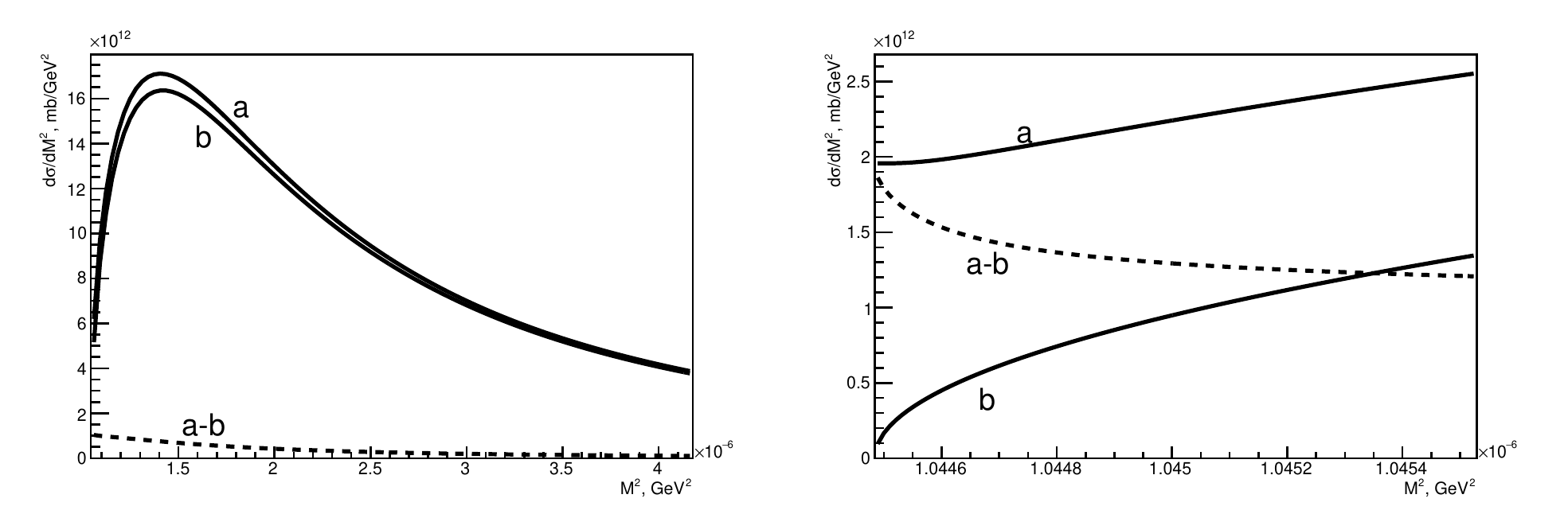}

Fig. 1. The distribution of masses of dielectrons produced in
ultraperipheral collisions at NICA energy $\sqrt {s_{nn}}$=11 GeV
with (a) and without (b) account of the SGS-factor. Their difference
(a-b) is shown by the dashed line. The region of small masses is shown
in the righthand side at the enlarged scale.
\end{figure}
\begin{figure}

\centerline{\includegraphics[width=\textwidth, height=8cm]{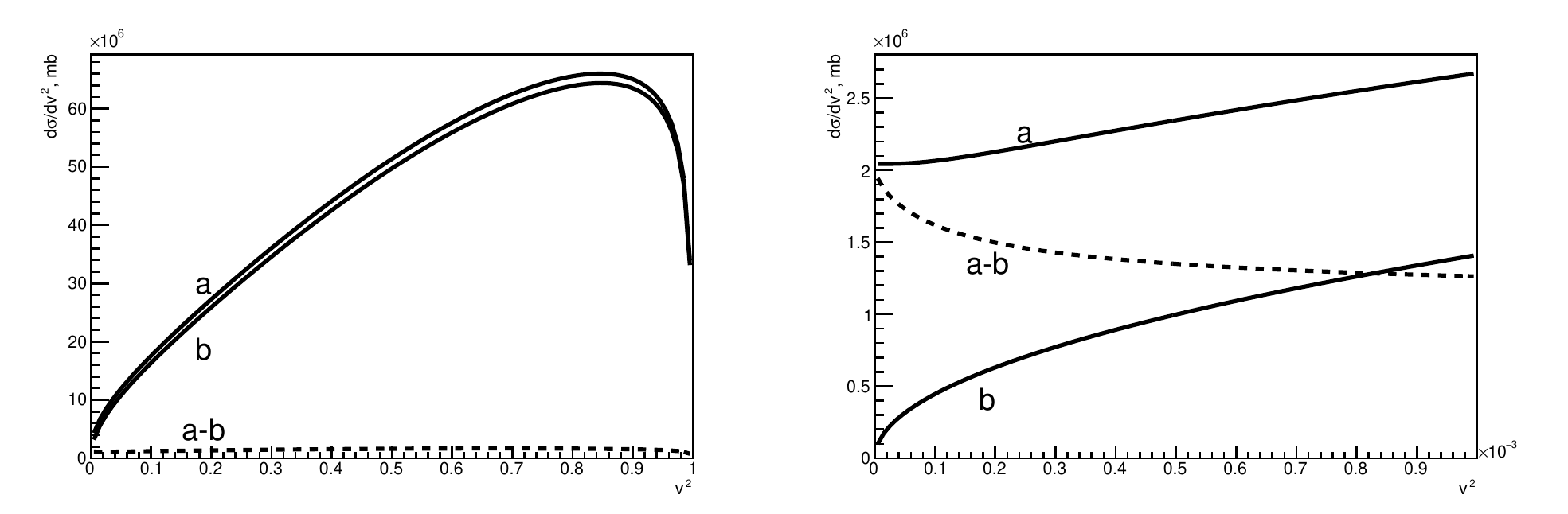}}

Fig. 2. The distribution of the relative velocities in dielectrons produced in
ultraperipheral collisions at NICA energy $\sqrt {s_{nn}}$=11 GeV
with (a) and without (b) account of the SGS-factor. Their difference
(a-b) is shown by the dashed line. The region of small masses is shown
in the righthand side at the enlarged scale.
\end{figure}

The cross sections of ultraperipheral production of $e^+e^-$-pairs are
especially strongly enhanced at low masses $M$ (at low relative velocities $v$)
compared to their perturbative values (marked by b). It is clearly seen
in the righthand sides of Figs 1 and 2 which demonstrate the region
near the threshold $M=2m$. Surely, the cross section would tend to zero
at the threshold $M=2m$ due to the energy-momentum conservation laws
not fully respected by the simplified SGS-recipe. However, it must happen
in the tiny region near the threshold and can be neglected in integral
estimates.

The situation with production of muon pairs is similar. We demonstrate it in
Fig. 3 by plotting the distribution of masses of muon pairs produced in
ultraperipheral collisions at LHC energy $\sqrt {s_{nn}}$=5.02 TeV.

There is no principal difference between the general shapes in
Fig. 3 and Figs. 1 and 2. The scales on both axes are just changed.
The main result about the enhanced low-mass pairs persists at higher
energies as well.

\begin{figure}
\includegraphics[width=\textwidth, height=8cm]{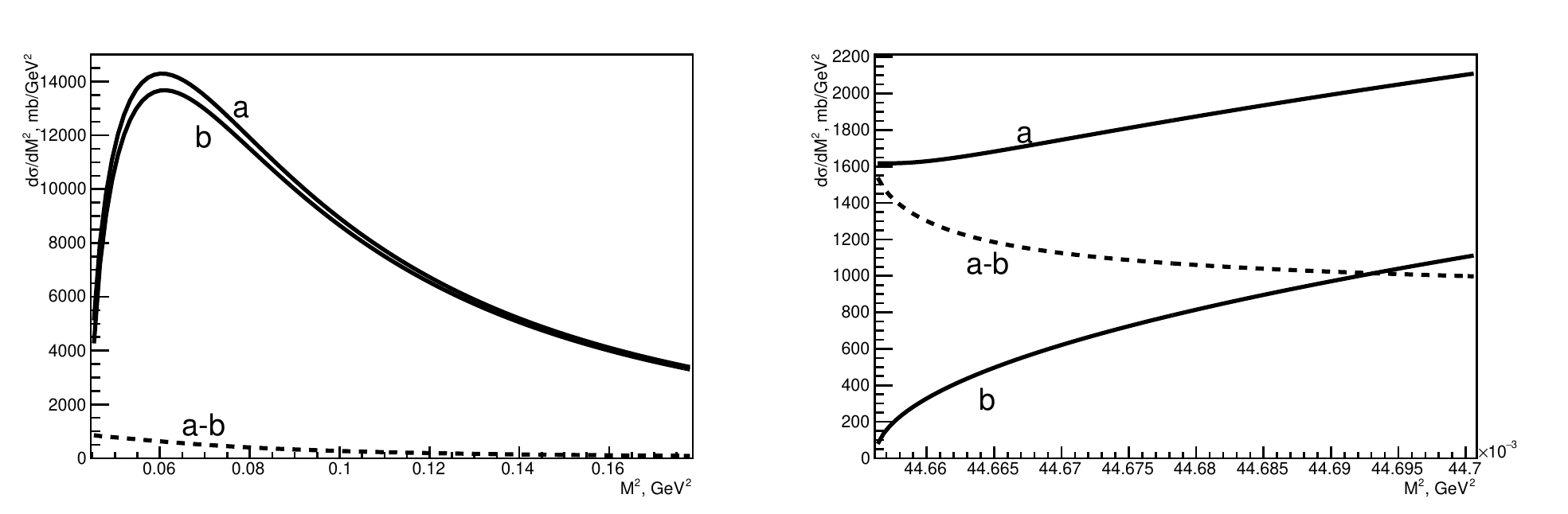}

Fig. 3. The distribution of masses of muon pairs produced in
ultraperipheral collisions at LHC energy $\sqrt {s_{nn}}$=5.02 TeV
with (a) and without (b) account of the SGS-factor. Their difference
(a-b) is shown by the dashed line. The region of small masses is shown
in the righthand side at the enlarged scale.
\end{figure}

The $e^+e^-$-pairs with low masses would be of no special interest if their
final products are electrons and positrons. The background is such high that
they are buried among many charged products of particle collisions.
However the mutual attraction of these components can result in their
annihilation. Two $\gamma $-quanta are created with definite energy $M/2$
in the rest system of a pair. This is the clear signature for their detection.
The energy distribution of quanta is peaked at 511 keV in NICA experiments as
shown in Ref.\cite{20dr} for the parapositronia decays. The observation of such
quanta can signal the onset of their ultraperipheral production there.

However, it is not excluded that the pairs with masses somewhat exceeding
the threshold $M=2m$ can also annihilate to two quanta with approximately the
same energy. The peaks will be slightly widened. The fate of the low-mass pairs
with slow relative motion of its components is hard to predict. The strong
Coulomb attraction is crucial.

The SGS-effect arises when an attractive interaction between
the non-relativistic particles significantly distorts their wave function,
such that they have a larger probability to undergo annihilation.
It has been claimed in Ref. \cite{ieng} that the quantum field theory
Bethe-Salpeter equation in the ladder approximation provides the leading
non-perturbative solution, with perturbative corrections coming from higher
order diagrams.
The non-relativistic nature of the pair of annihilating
particles separates the short-distance annihilation process (taking place at
 distances up to O(1/$m$)) from the long-distance interactions characterized by
the Bohr radius of the pair, responsible for the SGS-effect. The transition
is determined by their ratio which plays a role of the small parameter.

An anomalous low-mass dilepton excess demonstrated above in Figs 1 to 3
attracts much attention as a mechanism to boost the annihilation rates.
The ultraperipheral production of parapositronia is about $10^6$ times lower
than the supply of unbound pairs. The integral contribution of the low-mass
unbound pairs exceeds the parapositronia decay effect by several
orders of magnitude. 

The enhancement of low-mass $e^+e^-$-pairs can be extremely important for
understanding some observations of abundant production of the 511 keV
$\gamma $-quanta during the thunderstorms in the Earth atmosphere
\cite{abc} and a distinctive peak at this energy in signals from
the Universe \cite{sieg}. It was speculated in Ref. \cite{20dr} that
the common origin of these effects can be prescribed to the formation
of the dense electron-positron clouds by strong electromagnetic fields.
The widths of the peaks coming from different regions of the Milky Way
vary from 2.5 keV to 3.5 keV according to the results of Ref. \cite{sieg}.
These widths may correspond to the admissible intervals of masses $M$
ranging from $2m=1.022$ MeV to 1.027 MeV and 1.029 MeV, correspondigly,
and can be ascribed to annihilation of $e^+e^-$-pairs with such masses.
Integrating the plot a in Fig. 1 within these intervals of masses one gets
the cross sections equal to 40165 mb and 63296 mb which are much larger 
than the cross section of the parapositronium production about 20 mb. Therefore,
the outcome of $\gamma $-quanta with energies near 511 keV may become
more abundant than just from decays of parapositronia.
Studies at the NICA collider can help in getting the quantitative results
about properties of $e^+e^-$-pairs created in strong electromagnetic
fields of heavy ions during their ultraperipheral collisions.

\vspace{6pt}

{\bf Acknowledgements}

The work of I.D. was supported by the RFBR project 18-02-40131.

\end{document}